\documentstyle[12pt]{article}
\begin{document}
\begin{titlepage}

\title{Dirac spinor fields in the teleparallel gravity:
comment on ``Metric-affine approach to teleparallel gravity"}
\author{J. W. Maluf$\,^{*}$ \\
Instituto de F\'{\i}sica, \\
Universidade de Bras\'{\i}lia\\
C. P. 04385 \\
70.919-970 Bras\'{\i}lia DF\\
Brazil\\}
\date{}
\maketitle
\begin{abstract}

We show that the coupling of a Dirac spinor field $\psi$ with
the gravitational field in the teleparallel equivalent of general
relativity is consistent. For an arbitrary SO(3,1) connection
$\omega_{\mu ab}$ there are two possibilities for the coupling of
$\psi$ with the gravitational field. The problems of
consistency raised by Y. N. Obukhov and J. G. Pereira in the paper
{\it Metric-affine approach to teleparallel gravity} [gr-qc/0212080]
take place only in the framework of one particular coupling. By
adopting an alternative coupling the consistency problem disappears.

\end{abstract}
\thispagestyle{empty}
\vfill
\noindent PACS numbers: 04.20.Cv, 04.20.Fy, 04.90.+e\par
\bigskip
\noindent (*) e-mail: wadih@fis.unb.br\par
\end{titlepage}
\newpage

\noindent
The gravitational field can be described both by the usual
Riemannian geometry or by the teleparallel geometry. Although
the two geometrical frameworks are rather distinct,
it is well known that in each of them we may
obtain Einstein's equations provided we specify the appropriate
Lagrangian density\cite{Mol,Hay,Hehl,Nester1,Maluf1,Per}.
Therefore all the successful results of general relativity
that have been accomplished in the Riemannian geometry are also
achievements of the same theory in the teleparallel equivalent
description.

The teleparallel geometry is a suitable framework to address the
notions of energy, momentum and angular momentum of any space-time
that admits a 3+1 foliation\cite{Maluf2}. The principle of
equivalence is still subjacent to the idea, still endorsed by many
physicists, that the gravitational energy is not localizable.
However, it is unquestionable that the
results so far obtained in the realm of the teleparallel equivalent
of general relativity (TEGR) are relevant and deserve a throrough
and continued investigation.

The description of the gravitational field in the teleparallel
geometry is not unique. One can either establish a Lagrangian density
that is invariant under the global\cite{Mol,Hay,Nester1,Per,Maluf2} or
the local\cite{Hehl,YO,BL} SO(3,1) Lorentz group. Alternatively, we
may adopt tetrad fields $e^a\, _\mu$ that either transform under the
global or local Lorentz group (Notation: space-time indices $\mu, \nu,
...$ and SO(3,1) indices $a, b, ...$ run from 0 to 3. Time and space
indices are indicated according to $\mu=0,i,\;\;a=(0),(i)$. The flat,
Minkowski space-time  metric is fixed by
$\eta_{ab}=e_{a\mu} e_{b\nu}g^{\mu\nu}= (-+++)$). 
In the context of metric-affine
gravitational theories (MAG) it is natural to require the local
symmetry for the Lagrangian density. If the latter symmetry is
required, then one has to introduce a set of flat connection fields
$\omega_{\mu ab}=-\omega_{\mu ba}$. The zero curvature condition
$R_{ab\mu\nu}(\omega)=0$ is imposed by means of Lagrange multipliers
$\lambda^{ab\mu\nu}$ in the action integral. 

Therefore in empty space-time the TEGR can be described by the
Lagrangian density\cite{Hehl,YO,BL}

$$L_1(e,\omega,\lambda)\;=
\;-k\,e\,\biggl( {1\over 4} T^{abc}T_{abc} +
{1\over 2}T^{abc}T_{bac}-T^aT_a\biggr)+
\lambda^{ab\mu\nu}R_{ab\mu\nu}(\omega)\;,\eqno(1)$$

\noindent where $k=1/(16\pi G)$, $e=det(e^a\,_\mu)$,
$T_{abc}=e_b\,^\mu e_c\,^\nu T_{a \mu \nu}$ and
the trace of the torsion tensor is given by 
$T_b=T^a\,_{ab}\;.$ $L_1$ displays a local invariance under the
SO(3,1) group.
The torsion and curvature tensors are defined by

$$T^a\,_{\mu \nu}(e,\omega)
=\partial_\mu e^a\,_\nu-\partial_\nu e^a\,_\mu
+\omega_\mu\,^a\,_b\,e^b\,_\nu-
 \omega_\nu\,^a\,_b\,e^b\,_\mu\,,\eqno(2)$$

$$R^a\,_{b\mu\nu}=\partial_\mu \omega_\nu\,^a\, _b-
                 \partial_\nu \omega_\mu\,^a\, _b
+\omega_\mu\,^a\,_c\; \omega_\nu\,^c\,_b
-\omega_\nu\,^a\,_c\; \omega_\mu\,^c\,_b\,,\eqno(3)$$

\noindent respectively.

Alternatively, the TEGR can also be described by the following
Lagrangian density,

$$L_2(e)\;=
\;-k\,e\,\biggl( {1\over 4} T^{abc}T_{abc} +
{1\over 2}T^{abc}T_{bac}-T^aT_a\biggr)\;,\eqno(4)$$

\noindent where the definitions for $k$ and $e$ are the same as
in Eq. (1). However, the definition for $T^a\,_{\mu\nu}$ is now
simply given by

$$T^a\,_{\mu \nu}(e)=\partial_\mu e^a\,_\nu-\partial_\nu e^a\,_\mu
\;.\eqno(5)$$

The description of the gravitational field given by $L_2$ is
certainly simpler than the one given by $L_1$. Moreover, the theory
defined by $L_2$ is - {\it as it must be} - invariant under
coordinate transformations, a basic requirement of the general theory
of relativity. However, despite the rigid structure regarding the
frame invariance of $L_2$ (the symmetry group is the global SO(3,1)),
one can {\it choose any frame} in the description of {\it any}
solution of Einstein's equations. Any global frame yields a viable
teleparallel geometry of the space-time\cite{Nester2} and allows
the definition of distant parallelism, which is the prominent
feature of the teleparallel geometry.
In a space-time with an underlying tetrad field two  vectors at
distant points are called parallel\cite{Mol} if they have
identical components with respect to the local tetrads at
the points considered. Thus consider a vector field $V^\mu(x)$.
At the point $x^\lambda$ its tetrad components are given by
$V^a(x)=e^a\,_\mu(x)V^\mu(x)$. For the tetrad components
$V^a(x+dx)$ it is easy to show that $V^a(x+dx)=V^a(x)+DV^a(x)$,
where $DV^a(x)=e^a\,_\mu(\nabla_\lambda V^\mu)dx^\lambda$. 
The covariant derivative $\nabla$ is constructed out of 
Cartan's connection
$\Gamma^\lambda_{\mu \nu}=e^{a\lambda}\partial_\mu e_{a\nu}$
(the torsion tensor given by Eq. (5) is related to the torsion
tensor of Cartan's connection).
Therefore the vanishing of such covariant derivative defines
a condition for absolute parallelism in space-time.

In the interesting analysis given in Ref. \cite{YO} it is shown that
the Lagrange multipliers are determined from the field equations only
up to an ambiguity. In this framework there are field equations not
only for $e_{a\mu}$, but also for $\omega_{\mu ab}$ and 
$\lambda^{ab\mu\nu}$. By working out this larger set of field
equations one arrives at an inconsistency in the coupling of a Dirac
spinor field with the gravitational field\cite{YO}. It is shown in
this latter reference that one can only consistently couple the
gravitational field to a spinless matter or to a matter with a
conserved spin tensor.
This is a quite strong statement. We will show that if the
coupling is suitably defined, the interaction of $e_{a\mu}$ with
the Dirac field can be consistently carried out. 

Let us consider a Dirac spinor field $\psi(x)$ and a set of Dirac
matrices $\gamma^a$ that satisfy $\gamma^a\gamma^b+\gamma^b\gamma^a=
-2\eta^{ab}$. For an appropriate representation of $\gamma^a$ the
quantities $\Sigma^{ab}=i/2\lbrack \gamma^a,\gamma^b\rbrack$ are the
generators of the SO(3,1) group. We will consider initially an
arbitrary SO(3,1) connection $\omega_{\mu ab}$, in the spirit of
the MAG approach. Out of $\omega_{\mu ab}$ we construct the
derivative $D_\mu(\omega)$,
covariant under local SO(3,1) transformations,

$$D_\mu(\omega)\psi=\partial_\mu \psi -{i\over 4}\;\omega_{\mu ab}
\,\Sigma^{ab}\,\psi\;.\eqno(6a)$$

\noindent Similarly for $\bar \psi
= \psi^{\dagger} \gamma^{(0)}$ we have

$$D_\mu(\omega)\bar \psi=\partial_\mu \bar \psi+
{i\over 4}\;\omega_{\mu ab}\,\bar \psi \Sigma^{ab}\;.\eqno(6b)$$

With the help of Eqs. (6a,b) we establish the real valued
Lagrangian density $L_{M1}$ for a Dirac field of mass $m$,

$$L_{M1}(\psi,\bar \psi,\omega)
=e\biggl[ {i\over 2}\biggl(\bar \psi \gamma^\mu D_\mu(\omega)\psi
- (D_\mu(\omega) \bar \psi)\gamma^\mu \psi \biggr) -m \bar \psi \psi
\biggr]\;,\eqno(7)$$

\noindent where $\gamma^\mu(x)=e_a\,^\mu \gamma^a$. Variation of
$L_{M1}$ with respect to $\psi$ yields the Dirac equation,

$$i\gamma^\mu D_\mu(\omega) \psi-m \psi +
{i\over 2}e^{a\lambda}T_{a\mu\lambda}(e,\omega)\gamma^\mu \psi
=0\;.\eqno(8)$$

The theory considered by Obukhov and Pereira\cite{YO} is equivalent
to the one defined by the Lagrangian density $L=L_1+L_{M1}$. In
the framework of this latter Lagrangian there arises the consistency
problem indicated in Ref. \cite{YO}. However, a different approach
to the coupling of $\psi$ with $e^a\, _\mu$ is possible.

Let us consider the SO(3,1) connection $^o\omega_{\mu ab}$ given by

$$^o\omega_{\mu ab}=-{1\over 2}e^c\,_\mu(
\Omega_{abc}-\Omega_{bac}-\Omega_{cab})\;,\eqno(9)$$

$$\Omega_{abc}=e_{a\nu}(e_b\,^\mu\partial_\mu e_c\,^\nu-
                        e_c\,^\mu\partial_\mu e_b\,^\nu)\;,$$

\noindent This connection allows the definition of the SO(3,1)
covariant derivative

$$D_\mu(^o\omega)
\psi = \partial_\mu \psi -{i\over 4}\;^o\omega_{\mu ab}
\,\Sigma^{ab}\,\psi\;.\eqno(10a)$$

\noindent Likewise, for $\bar \psi
= \psi^{\dagger} \gamma^{(0)}$ we have

$$D_\mu(^o\omega) \bar \psi=\partial_\mu \bar \psi+
{i\over 4}\;^o\omega_{\mu ab}\,\bar \psi \Sigma^{ab}\;.\eqno(10b)$$

\noindent In similarity to $L_{M1}$ we construct the Lagrangian
density

$$L_{M2}(\psi,\bar \psi,\;^o\omega)
=e\biggl[ {i\over 2}\biggl(\bar \psi \gamma^\mu D_\mu(^o\omega)\psi
-(D_\mu(^o\omega)\bar \psi)\gamma^\mu \psi \biggr) -m \bar \psi \psi
\biggr]\;.\eqno(11)$$

The Lagrangian density that we will adopt in the framework of the
TEGR is similar to Eq. (11). In order to verify this issue, let us
consider a totally arbitrary SO(3,1) connection $\omega_{\mu ab}$.
It is well known that this connection can be identically written as

$$\omega_{\mu ab}=\;^o\omega_{\mu ab}+K_{\mu ab}\,,\eqno(12)$$

\noindent where $K_{\mu ab}$ is the usual contorsion tensor,

$$K_{\mu ab}={1\over 2}e_a\,^\lambda\,e_b\,^\nu(
T_{\lambda \mu \nu}+T_{\nu \lambda \mu}
-T_{\mu \nu\lambda})\;.$$

\noindent The torsion tensor in this expression is given
by (2).

In Eq. (12) both $\omega_{\mu ab}$ and $^o\omega_{\mu ab}$
transform as SO(3,1) connections and $K_{\mu ab}$ transforms as a
tensor. A vanishing connection $\omega_{\mu ab}=0$ is a possible
solution of the field equations that follow from $L_1$. Now, if
we require $\omega_{\mu ab}=0$ in Eq. (12), the local SO(3,1)
symmetry is reduced to the global SO(3,1) symmetry, and
$L_1(e,\omega,\lambda)$ reduces to $L_2(e)$. From Eq. (12) we
find that

$$^o\omega_{\mu ab}= -K_{\mu ab}\;,\eqno(13)$$

\noindent and in this case the covariant derivatives (10a,b) become

$$D_\mu(^o\omega)\psi = \partial_\mu \psi +{i\over 4}\;K_{\mu ab}
\,\Sigma^{ab}\,\psi\;,\eqno(14a)$$

$$D_\mu(^o\omega) \bar \psi=\partial_\mu \bar \psi-
{i\over 4}\;K_{\mu ab}\,\bar \psi \Sigma^{ab}\;,\eqno(14b)$$

\noindent which is precisely the covariant derivative presented
in Ref. \cite{VCA}. In Eqs. (13) and (14) the contorsion tensor
is defined in terms of the torsion tensor given by Eq. (5).

Therefore in the presence of Dirac spinor fields the total
Lagrangian density in the TEGR may be given by $L_2+L_{M2}$, where
$L_{M2}$ is constructed either with Eqs. (10a,b) or (14a,b).

It turns out that $L_T=L_2+L_{M2}$ yields a consistent set of
field equations. It has been shown by Coote and
Macfarlane\cite{CM} that the energy-momentum tensor that arises
from $L_T=keR(e,\omega)+L_{M2}$ is symmetric. More precisely, we
have

$$T_{\mu \nu}={1\over e} e_{a\nu}
{{\delta L_{M2}}\over {\delta e_a\, ^\mu}}=T_{\nu\mu}\;.\eqno(15)$$

\noindent Coote and Macfarlane have shown that anti-symmetric
components on the right hand side of the equations

$$R_{\lambda \mu}-{1\over 2} g_{\lambda \mu}R=
-{1\over {2k}} T_{\mu\nu}\;,\eqno(16)$$

\noindent vanish in virtue of the Dirac equation,

$$i\gamma^\mu D_\mu(^o\omega) \psi-m \psi=0\;,\eqno(17)$$

\noindent that is obtained from the action integral determined
by Eq. (11). Therefore the
anti-symmetric part of $T_{\mu\nu}$ does not pose any extra
conditions or restrictions on the field quantities. The resulting
field equations are just Einstein and Dirac equations.

In the framework of the TEGR a similar situation takes place.
Variation of $L_T=L_2+L_{M2}$ with respect to $e^{a\mu}$
yields\cite{Maluf1}

$${{\delta L_2}\over {\delta e^{a\mu}}}\;=\;
e_{a\lambda}e_{b\mu}\partial_\nu(e\Sigma^{b\lambda \nu})-
e\biggl(\Sigma^{b \nu}\,_aT_{b\nu \mu}-
{1\over 4}e_{a\mu}T_{bcd}\Sigma^{bcd}\biggr)
=-{1\over {4k}}e T_{a\mu}\;,\eqno(18)$$

\noindent where

$$\Sigma^{abc}=
{1\over 4}(T^{abc}+T^{bac}-T^{cab})+
{1\over 2}(\eta^{ac}T^b-\eta^{ab}T^c)\;.$$

\noindent In view of the identity

$${{\delta L_{2}}\over {\delta e^{a\mu}}}\; \equiv \;{1\over 2}\,e\,
\biggl\{ R_{a\mu}(e)-{1\over 2}e_{a\mu}R(e)\biggr\}\;,$$

\noindent we conclude that the field equations derived from
$L_T=L_2+L_{M2}$ are just Eqs. (16) and (17) only. Therefore the
Dirac spinor field can be consistently coupled to $e^a\,_\mu$.
It must be noted that the result obtained by Coote and Macfarlane
does not depend on the local SO(3,1) symmetry of the Lagrangian
density $L_T=keR(e,\omega)+L_{M2}$ or of the field equations.
Therefore it applies to $L_T=L_2+L_{M2}$ as well.

In conclusion, the total Lagrangian density defined by Eqs. (4) and
(11), the latter constructed out of the covariant derivatives
(14a,b), yields a consistent coupling of the Dirac and
gravitational fields.

\end{document}